\documentclass{aa}
\newcommand{\bit}{\begin{itemize}}
\newcommand{\eit}{\end{itemize}}
\newcommand{\ben}{\begin{enumerate}}
\newcommand{\een}{\end{enumerate}}
\newcommand{\bde}{\begin{description}}
\newcommand{\ede}{\end{description}}

\def\msol{{M}_{\odot}}

\def\grsim{\,\lower 1mm \hbox{\ueber{\sim}{>}}\,}
\def\lesssim{\,\lower 1mm \hbox{\ueber{\sim}{<}}\,}
\def\ueber#1#2{{\setbox0=\hbox{$#1$}
  \setbox1=\hbox to\wd0{\hss$ #2$\hss}
  \offinterlineskip
  \vbox{\box1\box0}}{}}
\def\degr{\hbox{$^\circ$}}
\def\arcmin{\hbox{$^\prime$}}
\def\arcsec{\hbox{$^{\prime\prime}$}}

\usepackage{graphicx}
\usepackage{multirow}
\usepackage{natbib}
\bibpunct{(}{)}{;}{a}{}{,}

\begin{document}

   \title{XMM observation of the dynamically young galaxy cluster CL~0939+4713}

   \author{E. De Filippis\inst{1,2},
           S. Schindler\inst{3,1},
           A. Castillo-Morales\inst{4,1}}

   \institute{Astrophysics Research Institute, 
               Liverpool John Moores University,  
               Birkenhead CH41 1LD,
               United Kingdom
\and 
	       Center for Space Research,
               Massachusetts Institute of Technology,
               70 Vassar Street, Building 37,
               Cambridge, MA 02139, USA
\and 
               Institut f\"ur Astrophysik,
               Leopold-Franzens-Universit\"at Innsbruck,
               Technikerstr. 25,
               A-6020 Innsbruck, Austria
\and
               Dpto.  F\'{\i}sica Te\'orica y del Cosmos, 
               Universidad de Granada, 
               Avda. Fuentenueva s/n, 18002 Granada, Spain}

\offprints{E. De Filippis,
\email{bdf@space.mit.edu}}

   \date{Received 13-08-02/ Accepted 24-03-03}

\abstract{
We present an XMM observation of the distant galaxy cluster CL~0939+4713. The X-ray image shows pronounced substructure, with two main subclusters which have even some internal structure. This is an indication that the cluster is a dynamically young system. This conclusion is supported by the temperature distribution: a hot region is found between the two main subclusters indicating that they are at the beginning of a major merger, and that they will probably collide in a few hundreds of Myr. The intra-cluster gas of CL~0939+4713 shows inhomogeneities in the metal distribution, with the optically richer subcluster having a higher metallicity.\\
}
\authorrunning {E. De Filippis et al.}
\titlerunning {XMM observation of CL~0939+4713}
   \maketitle

%
%________________________________________________________________

\section{Introduction}
The galaxy cluster CL~0939+4713 (Abell 851) at a redshift $z=0.41$ is an exceptionally rich cluster. It was extensively observed in morphological and spectroscopic studies of the galaxy population. These studies are based on ground-based optical and near-infrared~\citep{Dre92,Bel95,Fuk95,Sta95,Bel96,Dre99,Pog99,Iye00,Mar00} and HST observations~\citep{Dre93,Dre94,Sma99,Zie99,Fer00} as well as observations in UV~\citep{Bus00} and sub-millimeter~\citep{Cow02} wavelengths. A high number of star-burst and post-starburst galaxies was found in this distant cluster. To study the evolution of galaxies this number was compared with the corresponding number in nearby clusters. In the surroundings of the clusters a network of galaxy filaments and subclumps was found~\citep{Kod01}. \\
The cluster was also the target of several gravitational lensing analyses~\citep{Sei96,Gei99,Iye00} aiming at the determination of the cluster  mass distribution. The comparison between the  masses determined by the weak lensing analysis by~\cite{Sei96} and the X-ray analysis by~\cite{Sch98} showed a discrepancy of a factor of 2-3.\\
Previous X-ray observations by ROSAT and ASCA showed that the cluster has a very irregular structure indicating that the cluster is not relaxed but dynamically young~\citep{Sch96a,Sch98}: this result is in good agreement with the conclusion from larger scales by~\cite{Kod01}. The formation and evolution of clusters depends sensitively on cosmological parameters like the mean matter density in the universe $\Omega_m$ \citep{Tho98,Jen98,Bei01}. Therefore it is important to determine the dynamical state of clusters at different redshifts, i.e. at different evolutionary states. The X-ray morphology alone is not the best indicator of the dynamical state, but it should be complemented with all other information available, e.g. the temperature map or the galaxy distribution. The combination of all findings gives a detailed picture of the state of a cluster. We do this analysis here for the moderately distant cluster CL~0939+4713, which can be used as a basis for comparisons at lower and higher redshifts.\\
The capability of XMM to perform spatially resolved spectroscopy can be used also to determine the distribution of the metal abundances. Not only the overall value of metallicity but also its spatial distribution gives important indications on the metal enrichment processes and the cluster formation process. 
Early enrichment would point preferentially towards galactic winds driven by supernovae~\citep{DeY78}. For this process one would expect a priori no metallicity variations, because there was enough time to distribute the metals uniformly throughout the cluster. However, numerical simulations of this mechanism gave very discordant results. \cite{agu01} and  \cite{Met94,Met97} concluded that winds are very efficient in enriching the ICM, but the latter authors found a very steep radial metallicity gradient, which is not observed in any cluster. \cite{Mur99} and  \cite{Gne98} on the other hand found that galactic winds play only a minor role.\\
At a later stage ram pressure stripping \citep{Gun72} is expected to become increasingly important. If considerable amounts of metals are stripped recently a spatial variation -- not necessarily radial gradients -- should be observable. To test these  effects it is particularly important to analyse clusters at different redshifts to see how the efficiency of the different enrichment mechanisms changes with time.

Throughout this paper we use $H_0=50\ \mathrm{km\ s}^{-1} \mathrm{Mpc}^{-1}$ and $q_0=0.5$; all errors are $90\%$ confidence levels.
%------------------------------------------------------------------------------
\section{Observation and data analysis}

\subsection{Observation and data processing}
The XMM-Newton observation of the galaxy cluster CL~0939+4713 (ID 0106460101) was performed with the European Photon Imaging Camera (EPIC) on November 6th, 2000 during the satellite's $167^{\rm th}$ revolution. It consists of one pointing obtained operating the EPIC cameras in the standard Full Frame Mode. For all the three EPIC cameras (MOS1, MOS2, and pn) the thin filter was used for a total exposure time of $51\ {\rm ks}$. \\
We used the SASv5.0.1 processing tasks {\it emchain} and {\it epchain} to generate calibrated event files from raw data.\\
The XMM-Newton observations are strongly affected by high internal and external background levels. Efficient ways of subtracting these background components were extensively discussed by~\cite{Pra01} and \cite{Lum02}.\\

\subsection{Flare rejection}
\begin{figure}
\centering
\includegraphics[angle=-90,width=8.8cm]{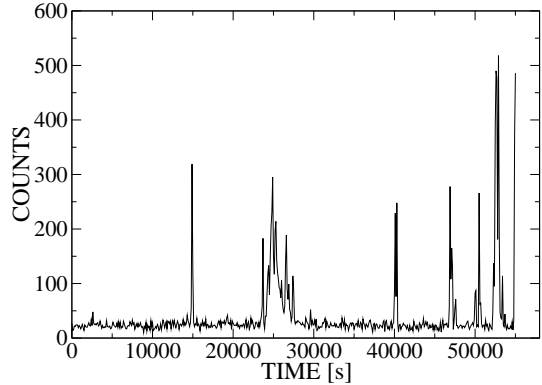}
\caption{MOS1 light curve in the energy range $10-12\ {\rm keV}$.}
\label{fig:MOS1_rates}
\end{figure}
Low energy ($< 1\ {\rm MeV}$) protons from solar flares are stopped in the first microns of silicon of the EPIC CCDs. They give a signal identical to X-rays, which makes it impossible to recognize them via a pattern selection. We obtained a cleaner set of data by visually inspecting the light curves and applying an intensity filter. In the energy range $10-12\ {\rm keV}$ for MOS1 \& MOS2 cameras ($12-14\ {\rm keV}$ for pn) we excluded all the intervals of exposure time having a count rate higher than a certain threshold value.\\
\begin{figure}
\centering
\includegraphics[width=8.8cm]{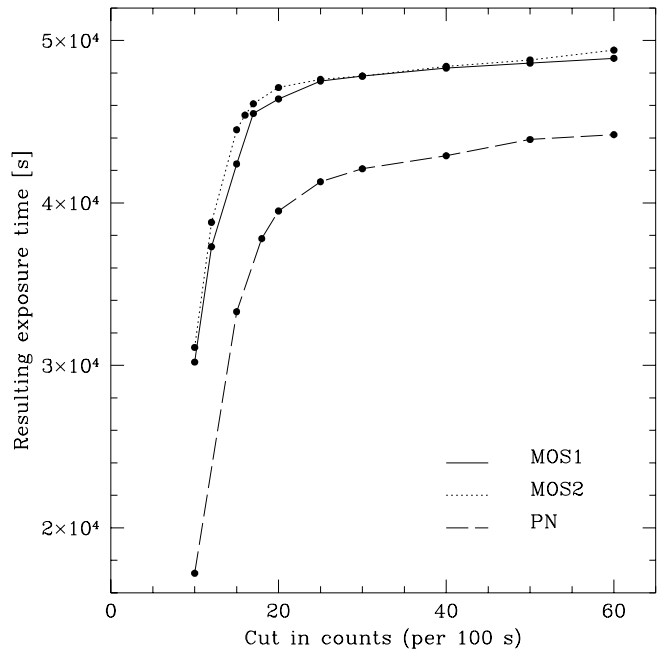}
\caption{Effective exposure time versus the threshold in the count rate in the energy band $10-12\ {\rm keV}$ and $12-14\ {\rm keV}$ for MOS1 \& MOS2 and for pn, respectively.}
\label{fig:exp_vs_counts}
\end{figure}

\begin{figure*}
\centering
\includegraphics[width=8.8cm]{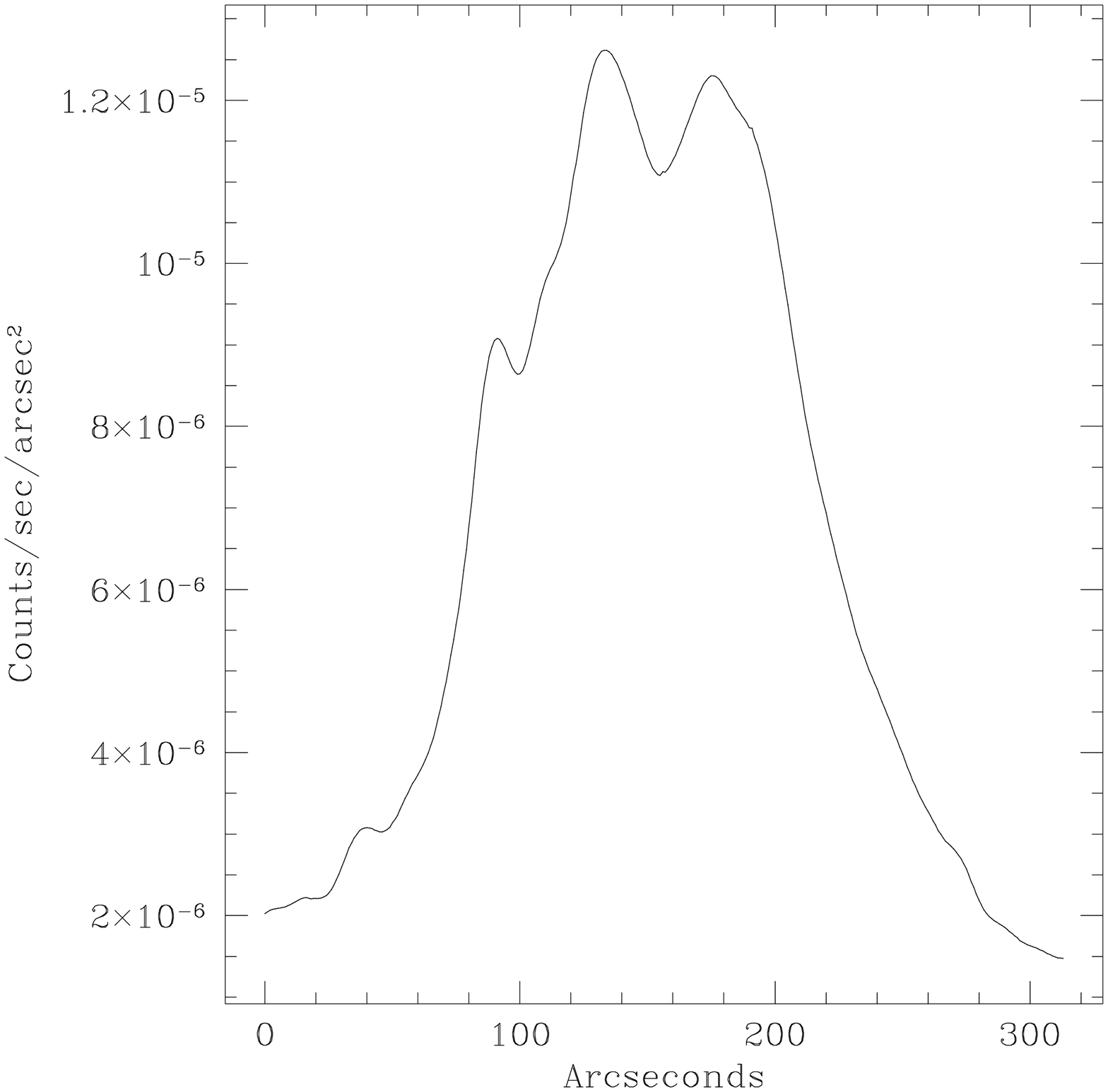}
\includegraphics[width=8.8cm]{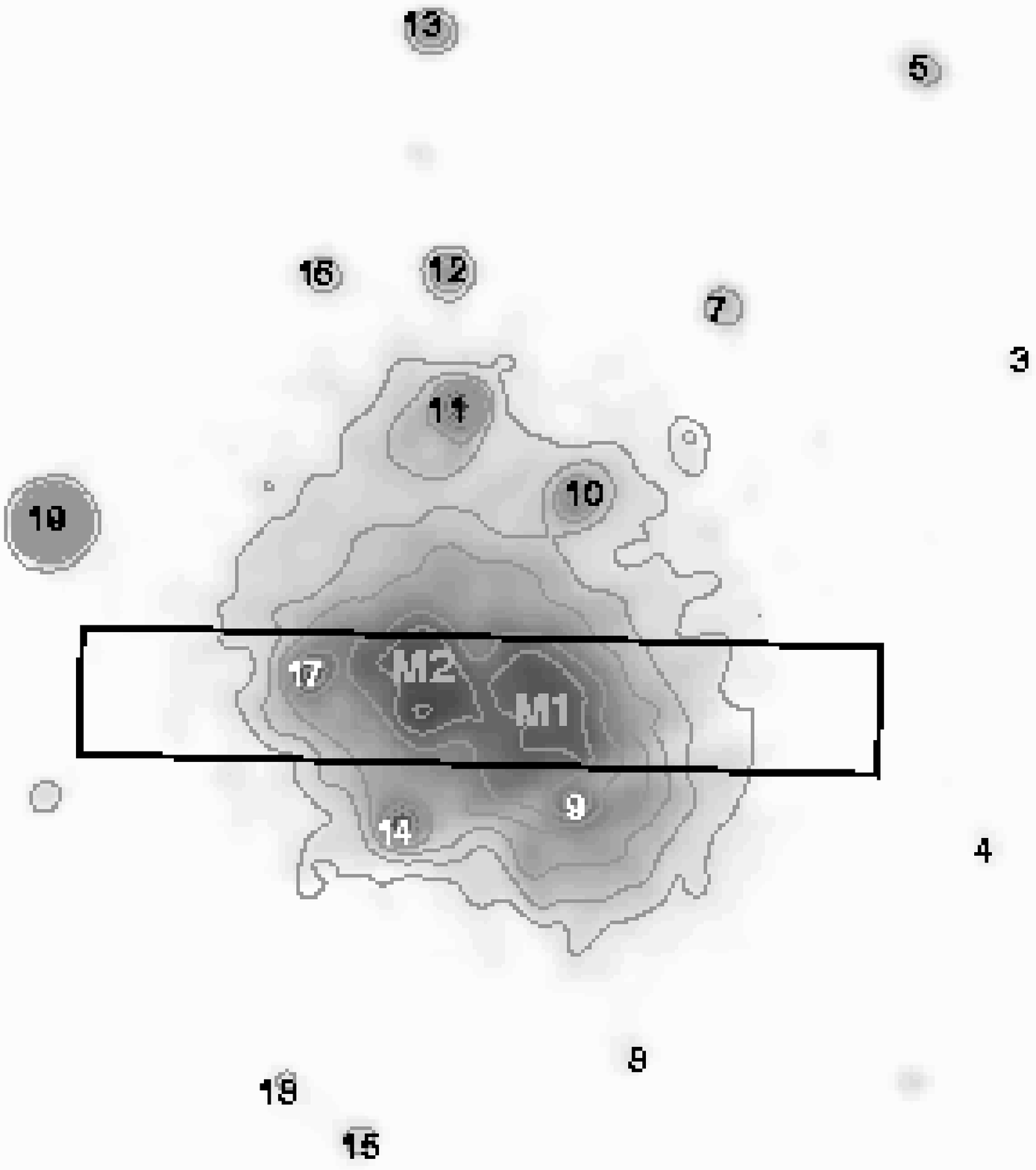}
\caption{{\bf Left}: Projection of the cluster surface brightness extracted from the rectangular region shown in the right panel. {\bf Right}: X-ray image of the cluster. Superposed are the linearly spaced X-ray contours, the identification numbers of the point sources detected in the field (see \S~\ref{sec:point}) and the location of the two subclusters forming the cluster core: M1 and M2, correspondent to the two maxima in the X-ray emission.}
\label{fig:projection}
\end{figure*}
The three light curves, one for each instrument, were analyzed separately in order to find the best threshold values (see Fig.~\ref{fig:MOS1_rates}). Different threshold values lead to different cuts in the exposure time (see Fig.~\ref{fig:exp_vs_counts}); the selected values were therefore a compromise between maximizing the exposure time and minimizing the presence of events caused by flares.\\
The chosen threshold values are 15 counts $\cdot 100\ {\rm s}^{-1}$ for MOS1 and MOS2 and 25 counts $\cdot 100\ {\rm s}^{-1}$ for pn; the threshold for the pn camera is higher because of the larger effective area of the instrument compared to MOS. The resulting effective exposure times for the three cameras are
\begin{center}
MOS1: $\ \ 45800\ {\rm s}$\\
MOS2: $\ \ 46300\ {\rm s}$\\
pn: $\ \ 41800\ {\rm s}$\\
\end{center}

\subsection{Vignetting correction}
\label{sec:vignet}
To correct for vignetting the method described in detail by~\cite{Arn01} has been used. A weighting, given by the ratio of the effective area at the centre of the detector -$A_{0,0}(E_i)$- to the effective area at the event position -$A_{x_j,y_j}(E_i)$-, is assigned to each photon extracted in each spectrum. The vignetting corrected photon counts -$C(E_i)$- in each energy channel $E_j$ ($E_i-\Delta E_i/2<E_j<E_i+\Delta E_i/2$) are
$$C(E_i)=\sum_j w_j (x_j,y_j)$$
with the weights $w_j$ being, as described above:
$$w_j=\frac{A_{0,0}(E_i)}{A_{x_j,y_j}(E_i)}$$
which therefore all have values greater than unity.\\
The errors in the spectra are recomputed taking into account the vignetting correction; the variances in each energy channel are therefore given by
$$\sigma^2(C(E_i))=\sum_j w_j^2.$$
The above method is easily generalized for imaging, as described by~\cite{Maj02}.

\subsection{Background subtraction}
\label{sec:back}
In order to subtract both the cosmic and the non cosmic X-ray background~\citep{Arn02}, the blank field observations, collected by D. Lumb were used.~\footnote{Event files can be downloaded via anonymous ftp at {\it xmm.vilspa.esa.es} in the directory /pub/ccf/constituents/extras/background.} These are homogeneous collections of high Galactic latitude observations where no bright source is present, for a total exposure time of about $400\ {\rm ksec}$. These observations were all taken in a period of very low flare occurrence and almost exclusively with the use of a thin filter; bad pixels and sources were also removed.\\
We applied to the background event files the same intensity filter applied to our observations; this yields final exposure times for the blank fields of
MOS1: $\ \ 373425\ {\rm s}$,
MOS2: $\ \ 358904\ {\rm s}$, and
pn: $\ \ 332646\ {\rm s}$. 
Since the cosmic ray induced background might slightly change with time, the count rate in the $10.0-12.0\: {\rm keV}$ ($12.0-14.0\: {\rm keV}$) energy band for MOS1 \& MOS2 (pn) cameras in the blank field was compared with the count rate in our observation (after flare rejection). The obtained normalization factors ($0.995$, $0.992$ and $1.038$ for MOS1, MOS2 and pn, respectively) are then used to renormalize the blank field data.\\
We then looked for a possible residual soft background due to the soft component (${\rm E}<1.5\ {\rm keV}$) of the cosmic X-ray diffuse emission, which varies with the position in the sky. At the position in the sky of CL~0939+4713, no excess of soft cosmic background was observed in previous observations~\citep{Sno97}.\\
Nevertheless, we performed a further check. We chose a region in our pointing where no emission from the cluster, or from other point sources, was observed. For all the three detectors, spectra were extracted from this region, together with background spectra from the blank fields by D. Lumb. No evident excess of flux, associated with the soft X-ray background, is observed in any of the three detectors. We then subtracted the blank field spectra from the ones extracted in our observation; the resulting spectra yield no significant signal and are dominated by noise only.\\
No additional correction for excess of soft background is therefore needed for CL~0939+4713.\\

%__________________________________________________________________

\section{Morphological analysis}
Figure~\ref{fig:smoo_mosaic} shows an image of CL~0939+4713 in the $0.3-2.0\ {\rm keV}$ band. The image is an adaptively smoothed and exposure corrected mosaic of photon events from the three detectors (MOS1, MOS2 and pn).\\
The cluster emission has an elliptical shape, centred at RA$=9^{\rm h}\ 43^{\rm m}\ 01.00^{\rm s}$, DEC$=+46\degr\ 59\arcmin\ 37.50\arcsec$ (J2000). The axis ratio of major to minor axis is $1.3$. The position angle of the ellipse is P.A.$\approx 63\degr$ (N over E).\\

\subsection{Substructure}
\begin{figure*}
\centering
\includegraphics[width=17cm]{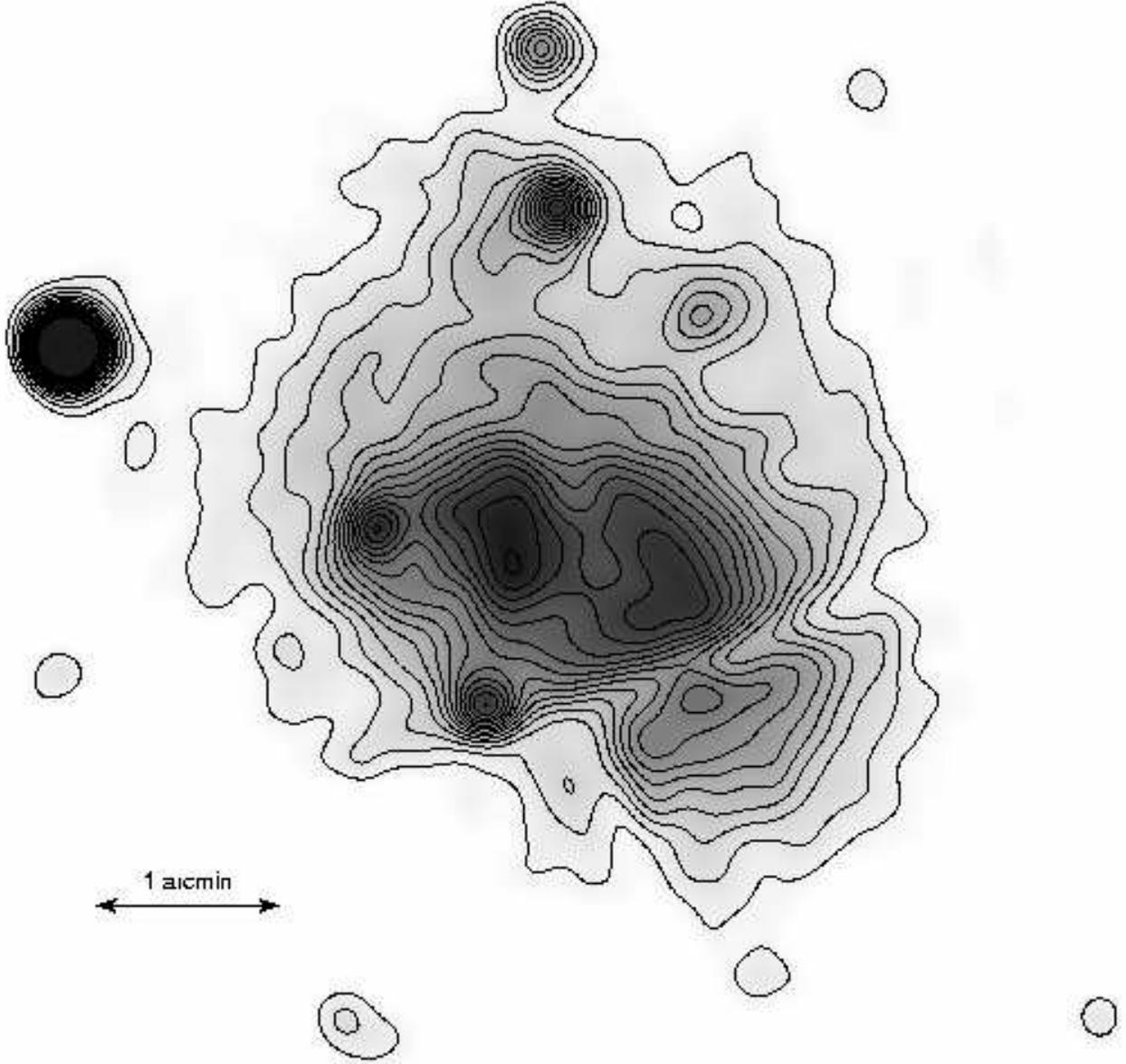}
\caption{X-ray image of the galaxy cluster CL~0939+4713 in the energy range 0.3-2.0 keV. The image is a mosaic of the three MOS1, MOS2 \& pn cameras. The core of CL~0939+4713 does not show a simple single central structure, but is composed of two subclusters. The two subclusters in turn show internal structure. Several other point sources are visible. The X-ray image is adaptively smoothed with a Gaussian with $\sigma_{\rm max}=$ 6 arcsec to keep an uniform signal-to-noise ratio of 3 $\sigma$. The contours are linearly spaced with a spacing of 0.006 counts s$^{-1}$ arcmin$^{-2}$. The highest contour line corresponding to 0.10 counts s$^{-1}$ arcmin$^{-2}$.}
\label{fig:smoo_mosaic}
\end{figure*}

Figure~\ref{fig:smoo_mosaic} clearly indicates that the core of CL~0939+4713 does not consist of a single spherically symmetric structure, but is composed of two subclusters: M1 and M2.\\
In Fig.~\ref{fig:projection} (left panel) a surface brightness projection of the cluster is shown, which is extracted from the region shown in the right panel. The two extended substructures forming the core are of similar size and of similar brightness, the subcluster in the East --M2-- being slightly brighter than the subcluster in the West --M1--). They both have an irregular shape, showing signs of further internal substructure. Their shapes are roughly elliptical with minor axis of $\approx 120\ {\rm kpc}$ ($19\arcsec$) and major axis of $\approx 180\ {\rm kpc}$ ($28\arcsec$).\footnote{$1\arcmin=389\ {\rm kpc}$ at the cluster distance of ${\rm z}=0.406$.} Their peaks are $\approx 280\ {\rm kpc}$ apart and their coordinates (J2000) are:
\begin{center}
${\rm P}_{\rm M1}$: RA$=9^{\rm h}\ 42^{\rm m}\ 58.9^{\rm s}$, DEC$=+46\degr\ 59\arcmin\ 36.0\arcsec$\\
${\rm P}_{\rm M2}$: RA$=9^{\rm h}\ 43^{\rm m}\ 03.3^{\rm s}$, DEC$=+46\degr\ 59\arcmin\ 26.5\arcsec$
\end{center}
They are slightly offsetted (${\rm P}_{\rm M1}$ $\approx 44\arcsec$ north and ${\rm P}_{\rm M2}$ $\approx 21\arcsec$ south) compared to the position observed by~\cite{Sch98}, and reach a maximum surface brightness of $5.0\times 10^{-2}\ {\rm counts\ s}^{-1}\ {\rm arcmin}^{-2}$ and $5.8\times 10^{-2}\ {\rm counts\ s}^{-1}\ {\rm arcmin}^{-2}$, respectively.\\
The smaller peak in Fig.~\ref{fig:projection} (left panel), located on the left of the projection at a distance of about $90\arcsec$ from the left border of the region, is caused by the presence of a QSO~\citep{Dre93} (see \S~\ref{sec:point}).\\

\subsection{Surface brightness profile}
\label{sect:surf}
\begin{figure*}
\centering
\includegraphics[width=7.8cm]{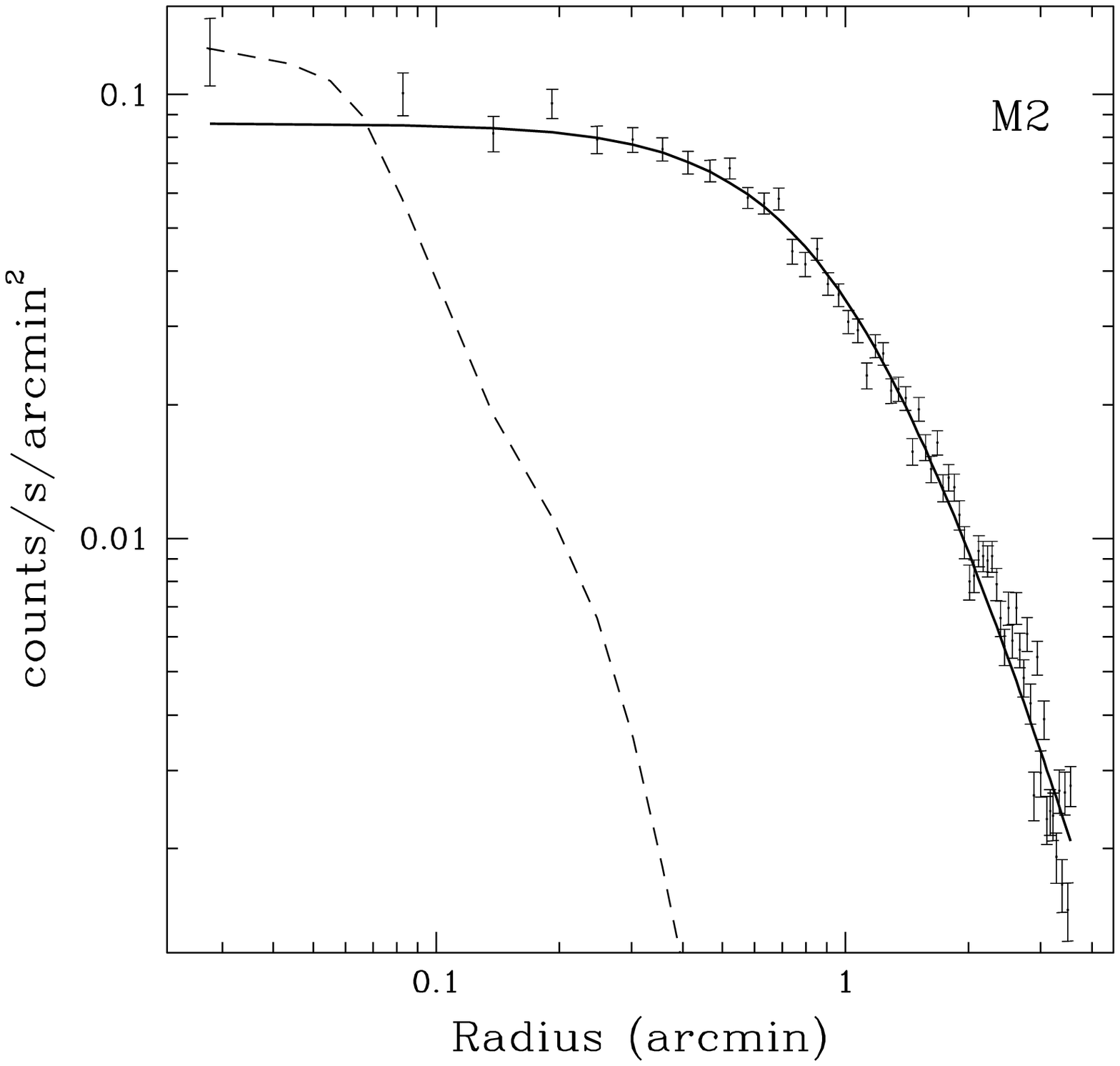}
\includegraphics[width=7.8cm]{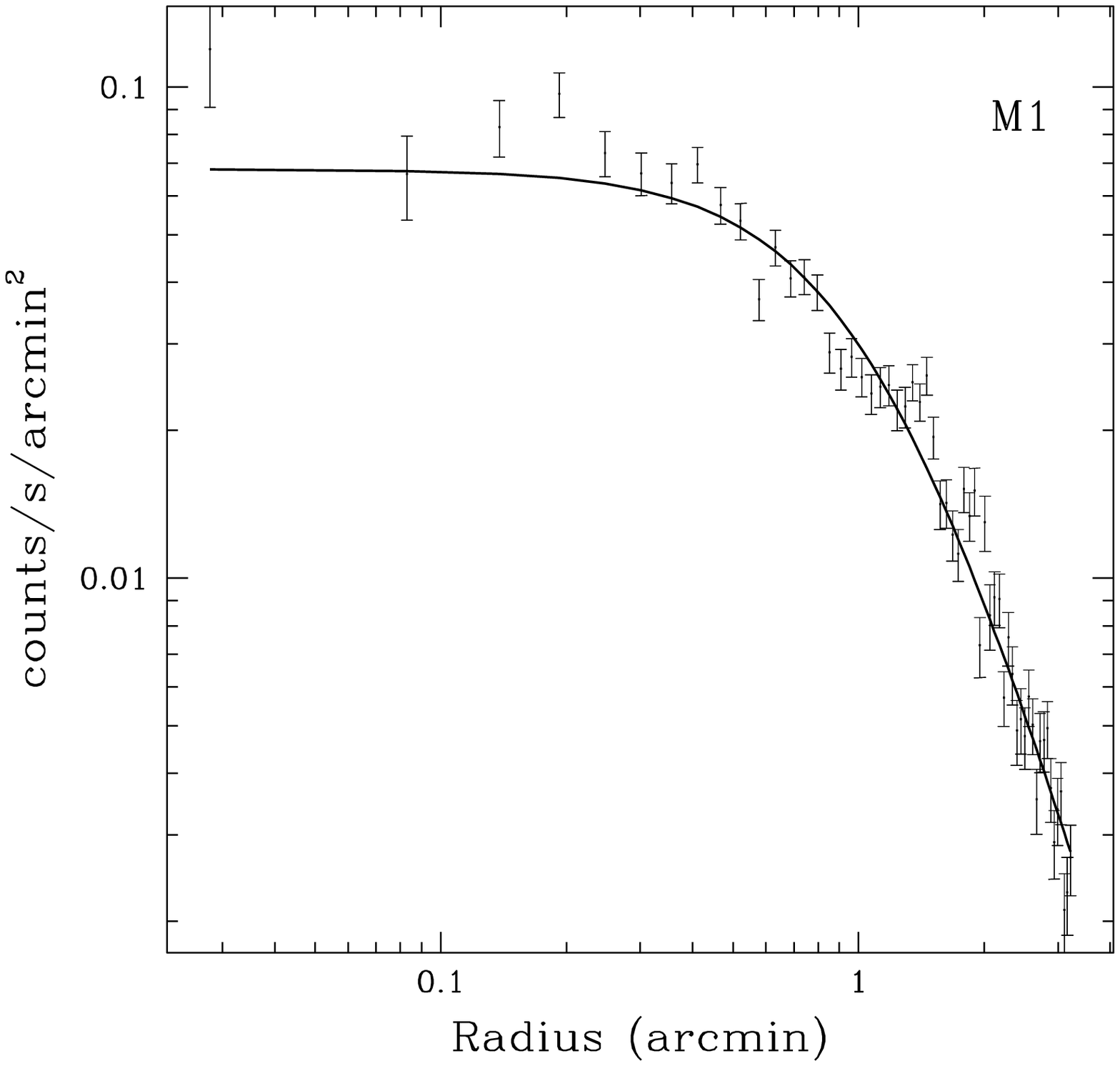}
\caption{Combined MOS1, MOS2 and pn surface brightness profile around the maxima M2 (left) and M1 (right) with spherically symmetric $\beta$-model fits (solid line). The fit parameters are given in Table~\ref{tab:parameters}. The dashed line shows the on-axis PSF, normalized to the central intensity.}
\label{fig:profM2M1}
\end{figure*}

\begin{figure}
\centering
\includegraphics[width=7.8cm]{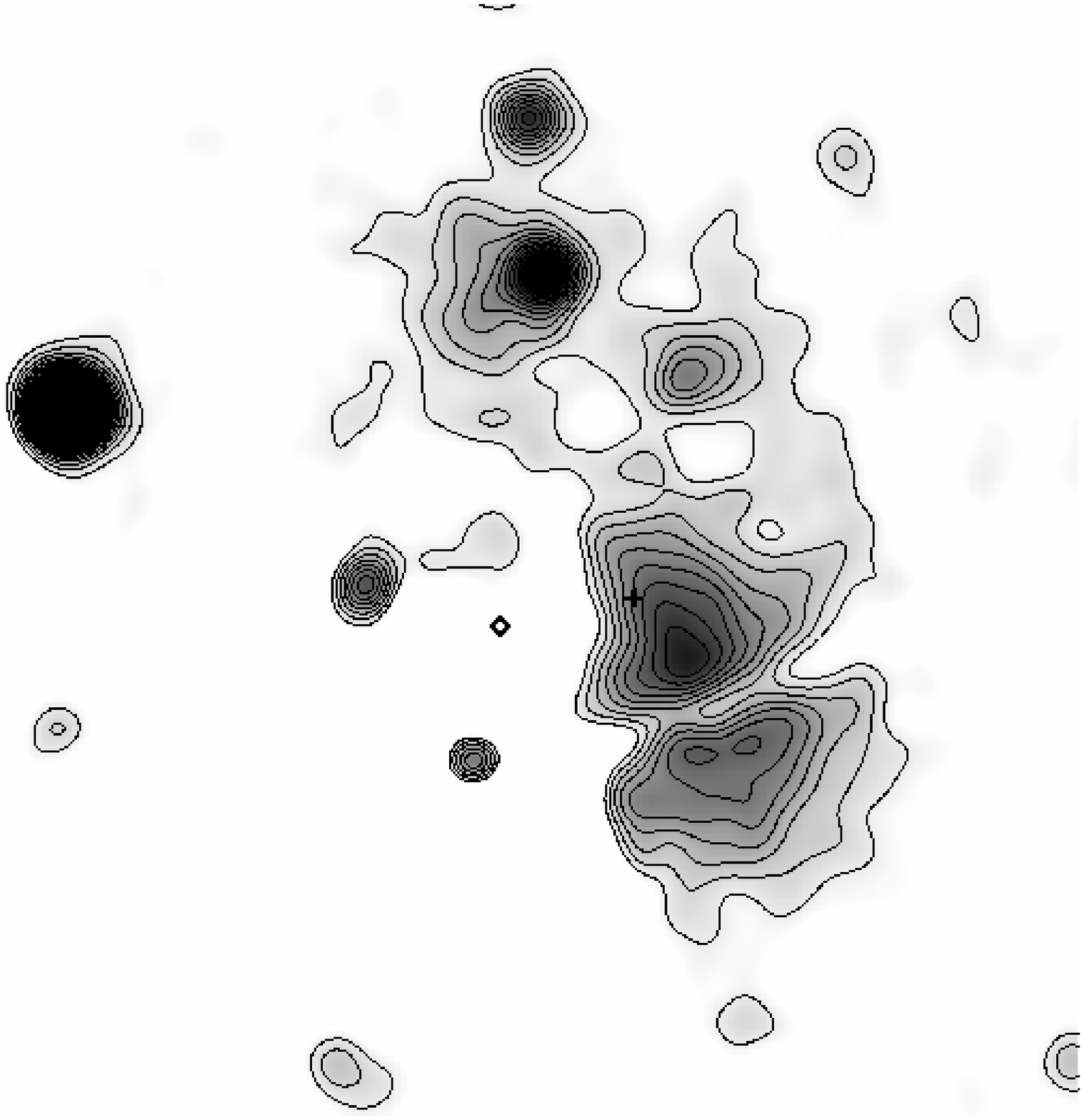}
\caption{Residual image in the energy band 0.3-2.0 keV after subtracting the spherical $\beta$ model centred on subcluster M2 from the original image. The positions of M1 (plus sign) and M2 (diamond point) are indicated. After the subtraction there is extended and structured emission around the centre M1. The emission from the quasar and the point sources that appear in Fig.~\ref{fig:smoo_mosaic} is visible. The contours are linearly spaced with a spacing of 0.005 counts s$^{-1}$ arcmin$^{-2}$. The highest contour line corresponding to 0.10 counts s$^{-1}$ arcmin$^{-2}$.}
\label{fig:residual}
\end{figure}

Although the cluster morphology is not spherically symmetric we perform a standard surface brightness analysis to obtain approximate values, which give a quantitative indication of the size of the cluster and also for comparison with other clusters. We generate an azimuthally averaged surface brightness profile for the cluster in the $0.3-2.0\ {\rm keV}$ energy band for each camera. This energy band was selected to optimize the signal$-$to$-$noise (S/N) ratio. Point sources were removed manually and the photons were binned into concentric annuli with a width of 3.3$\arcsec$ (3 pixels of the MOS camera). The vignetting correction and the background subtraction was done as described in \S~\ref{sec:vignet} and \S~\ref{sec:back}. The three profiles were then added.\\
As the cluster is clearly not spherically symmetric the profiles were centred on the positions of the two peaks (${\rm P}_{\rm M1}$ and ${\rm P}_{\rm M2}$) of both subclusters. The profile centred on ${\rm P}_{\rm M2}$ is obtained excluding a sector of 120$\degr$ in the direction of the other subcluster. In the case of the profile centred on ${\rm P}_{\rm M1}$ a wider sector was excluded (240$\degr$) in order to avoid the emission coming from the region located at the southern-west part of M1. The resulting surface brightness profiles $S(r)$ are shown in Fig.~\ref{fig:profM2M1} where the overall XMM point spread function (PSF) binned as the observed profiles is plotted as a dashed line. The overall PSF is obtained by adding the PSF of each camera, estimated at an energy of 1 keV and weighted by the respective cluster count rate in the $0.3-2.0\ {\rm keV}$ energy band. The comparison of the surface brightness profile data with the PSF shows that the effect of the PSF on the profile is negligible.\\
\begin{table*}[ht]
\caption[]{Fit parameters for a $\beta$ model centred on M1 and M2: $S_{0}$ is the central surface brightness, $r_{c}$ is the core radius, and $\beta$ is the slope parameter.}
\begin{center}
\begin{tabular}{l l l l }
\hline \hline
region& $S_{0}$ in counts s$^{-1}$ arcmin$^{-2}$& $r_{c}$ in arcmin& $\beta$ \cr
\hline
M2      &  8.6$\pm0.4\times10^{-2}$& 1.11$\pm0.07$ & 0.68$\pm0.03$\cr
M1      &  6.8$\pm0.5\times10^{-2}$& 1.16$^{+0.18}_{-0.15}$ & 0.66$^{+0.07}_{-0.05}$\cr
\hline
\end{tabular}
\end{center}
\label{tab:parameters}
\end{table*}
After the subtraction of the corresponding blank field profile we find no residual CXB background left (as already mentioned in Sec.~\ref{sec:back}). We therefore fit the surface brightness with a $\beta$-model without the need of an additional background component:
\begin{center}
$S(r)=S_{0}\left[1+\left(\frac{r}{r_{c}}\right)^{2}\right]^{-3\beta+1/2}$.
\end{center}
where $S_{0}$ is the central surface brightness, $r_{c}$ is the core radius, $\beta$ is the slope parameter. The profiles for the two subclusters M1 and M2 are fit out to radii of 3.5$'$ and 3.2$'$ respectively, at which the $S/N>3\sigma$. Results of the different fits are given in Table~\ref{tab:parameters}. We obtain similar fit parameters for the two subclusters; however the central region of the profile centred on ${\rm P}_{\rm M1}$ is not well fitted probably because of internal structure of the subcluster that flattens the central part. The derived beta and core radii parameters for both subclusters are larger compared with the ones obtained by~\cite{Sch98}. These discrepancies could be due to the low S/N ratio of the HRI observation that provides artificially low values for the beta-parameters. The HRI profiles decrease rapidly to background level (see Fig. 2 in~\cite{Sch98}), therefore a beta model fit is actually probing only the central part of the profile. In order to test this possibility decreasing cut off radii were applied to the XMM profiles. Smaller values of the core radius and beta were indeed derived: for a cut off radius of 2' $r_{c}=0.5'$ and $\beta=0.41$ were obtained for the profile centred on ${\rm P}_{\rm M1}$ and $r_{c}=0.8'$ and $\beta=0.57$ for the profile centred on ${\rm P}_{\rm M2}$.
For a more detailed analysis of the substructure, we subtract the emission coming from the subcluster M2 from the original image by using the $\beta$-model with the fit parameters in Table~\ref{tab:parameters}. The residual image is shown in Fig.~\ref{fig:residual}. In this residual image the emission from the quasar and from the point sources that appear in Fig.~\ref{fig:smoo_mosaic} is still visible. There is clearly extended emission left from the subcluster M1, but its peak has shifted to the south-west of about 25 arcsec. North of M1 there is also residual emission surrounding some point sources, which is clearly extended and structured. South-west of M1, separated by an artifact minimum caused by the gap in the detector chips is a last piece of residual extended emission. 
%-----------------------------

\subsection{Count rate}

The X-ray emission of the cluster can be traced out to $3.5$ arcminutes ($1.4\ {\rm  Mpc}$). The MOS and pn count rates within the $R\le3.5\arcmin$ region and in the considered energy band (0.3-2.0 keV) are $0.125\pm0.013$ counts s$^{-1}$ and $0.31\pm0.03$ counts s$^{-1}$ respectively. 

\subsection{Other sources in the pointing}
\label{sec:point}

In the pointing many other sources are found around the cluster. $19$ point sources, located in the area covered by MOS CCD$\#1$, were detected by both the pipeline detection and a peak density finding algorithm. Their coordinates, extension and count rates (after background subtraction) are listed in Table~\ref{tab:point_sou}. Most of these sources are shown in Fig.~\ref{fig:projection} (right panel) together with their identification number.\\
\begin{table*}[t]
\caption[]{List of point sources detected in the field covered by MOS CCD$\#1$. Columns 1-3 give the source identification number, and its coordinates (J2000). In Cols. 4 and 5 the source count rate (in counts per second) and the radius (in arcseconds) out to which the source can be detected, respectively, are listed.}
\begin{center}
\begin{tabular}{l l l l l}
\hline \hline
ID & RA & DEC & count rate  & radius  \cr
  &  &  &  & $\left [\arcsec \right]$ \cr
\hline
1 & $9^{\rm h}\ 42^{\rm m}\ 29.23^{\rm s}$ & $+46\degr\ 56\arcmin\ 25.27\arcsec$ & $0.0025$ & $30$ \cr
2 & $9^{\rm h}\ 42^{\rm m}\ 34.90^{\rm s}$ & $+46\degr\ 55\arcmin\ 58.35\arcsec$ & $0.0024$ & $30$ \cr
3 & $9^{\rm h}\ 42^{\rm m}\ 40.53^{\rm s}$ & $+47\degr\ 01\arcmin\ 40.41\arcsec$ & $0.0006$ & $22$ \cr
4 & $9^{\rm h}\ 42^{\rm m}\ 41.72^{\rm s}$ & $+46\degr\ 58\arcmin\ 32.43\arcsec$ & $0.0004$ & $20$ \cr
5 & $9^{\rm h}\ 42^{\rm m}\ 44.04^{\rm s}$ & $+47\degr\ 03\arcmin\ 36.45\arcsec$ & $0.0016$ & $25$ \cr
6 & $9^{\rm h}\ 42^{\rm m}\ 51.09^{\rm s}$ & $+47\degr\ 05\arcmin\ 14.49\arcsec$ & $0.0028$ & $35$ \cr
7 & $9^{\rm h}\ 42^{\rm m}\ 51.88^{\rm s}$ & $+47\degr\ 02\arcmin\ 04.49\arcsec$ & $0.0023$ & $25$ \cr
8 & $9^{\rm h}\ 42^{\rm m}\ 55.21^{\rm s}$ & $+46\degr\ 57\arcmin\ 10.50\arcsec$ & $0.0011$ & $17$ \cr
9 & $9^{\rm h}\ 42^{\rm m}\ 57.36^{\rm s}$ & $+46\degr\ 58\arcmin\ 46.50\arcsec$ & $0.0048$ & $15$ \cr
10 & $9^{\rm h}\ 42^{\rm m}\ 57.39^{\rm s}$ & $+47\degr\ 00\arcmin\ 50.12\arcsec$ & $0.0054$ & $22$ \cr
11 & $9^{\rm h}\ 43^{\rm m}\ 01.85^{\rm s}$ & $+47\degr\ 01\arcmin\ 24.50\arcsec$ & $0.0078$ & $30$ \cr
12 & $9^{\rm h}\ 43^{\rm m}\ 02.25^{\rm s}$ & $+47\degr\ 02\arcmin\ 16.50\arcsec$ & $0.0025$ & $22$ \cr
13 & $9^{\rm h}\ 43^{\rm m}\ 03.03^{\rm s}$ & $+47\degr\ 03\arcmin\ 52.50\arcsec$ & $0.0026$ & $25$ \cr
14 & $9^{\rm h}\ 43^{\rm m}\ 04.20^{\rm s}$ & $+46\degr\ 58\arcmin\ 40.49\arcsec$ & $0.0036$ & $15$ \cr
15 & $9^{\rm h}\ 43^{\rm m}\ 05.36^{\rm s}$ & $+46\degr\ 56\arcmin\ 36.49\arcsec$ & $0.0019$ & $18$ \cr
16 & $9^{\rm h}\ 43^{\rm m}\ 07.14^{\rm s}$ & $+47\degr\ 02\arcmin\ 16.48\arcsec$ & $0.0025$ & $25$ \cr
17 & $9^{\rm h}\ 43^{\rm m}\ 07.72^{\rm s}$ & $+46\degr\ 59\arcmin\ 40.48\arcsec$ & $0.0036$ & $15$ \cr
18 & $9^{\rm h}\ 43^{\rm m}\ 08.88^{\rm s}$ & $+46\degr\ 57\arcmin\ 00.45\arcsec$ & $0.0021$ & $27$ \cr
19 & $9^{\rm h}\ 43^{\rm m}\ 17.69^{\rm s}$ & $+47\degr\ 00\arcmin\ 38.40\arcsec$ & $0.0089$ & $30$ \cr
\hline
\end{tabular}
\end{center}
\label{tab:point_sou}
\end{table*}
For the sources corresponding to member galaxies, or located within $4.3\ {\rm arcmin}$ from the cluster centre, a more detailed analysis has been performed.
\begin{description}
\item[ID $\#3$] For this source no optical counterpart was found. The low statistics of the data do not allow us to constrain the photon index, even using a simple power law model. Assuming a power law spectrum with photon index $\Gamma=1.7$ we obtain an unabsorbed flux of $f_{\rm X}(0.3-10.0)=5.7\times 10^{-15}\ {\rm erg\ s}^{-1}{\rm cm}^{-2}$.
\item[ID $\#4$] No optical counterpart. Assuming a power law spectrum with photon index $\Gamma=1.7$: $f_{\rm X}(0.3-10.0)=3.5\times 10^{-15}\ {\rm erg\ s}^{-1}{\rm cm}^{-2}$.
\item[ID $\#7$] No optical counterpart. Assuming a power law spectrum with photon index $\Gamma=1.7$: $f_{\rm X}(0.3-10.0)=2.2\times 10^{-14}\ {\rm erg\ s}^{-1}{\rm cm}^{-2}$.
\item[ID $\#8$] Probable member galaxy Abell 0851:[DG92] 124. Assuming a power law spectrum with photon index $\Gamma=1.7$: $f_{\rm X}(0.3-10.0)=1.0\times 10^{-14}\ {\rm erg\ s}^{-1}{\rm cm}^{-2}$, and a bolometric luminosity of $L_{\rm X}{\rm (bol)}=1.6\times 10^{43}\ {\rm erg\ s}^{-1}$.
\item[ID $\#9$] Member galaxy Abell 0851:[DG92] 311 at a redshift of $z=0.4007$. It is also a radio source known as [CDS90] R082. Its spectrum was fitted with a power law giving a photon index $\Gamma=1.6\pm0.2$; the unabsorbed flux is $f_{\rm X}(0.3-10.0)=4.7\times 10^{-14}\ {\rm erg\ s}^{-1}{\rm cm}^{-2}$, and the bolometric luminosity $L_{\rm X}{\rm (bol)}=7.5\times 10^{43}\ {\rm erg\ s}^{-1}$.
\item[ID $\#10$] Probable member galaxy Abell 0851:[SED95] 074. The spectral fit yields $\Gamma=1.3\pm0.1$; $f_{\rm X}(0.3-10.0)=6.5\times 10^{-14}\ {\rm erg\ s}^{-1}{\rm cm}^{-2}$; $L_{\rm X}{\rm (bol)}=1.1\times 10^{44}\ {\rm erg\ s}^{-1}$.
\item[ID $\#11$] This source was already detected during a ROSAT PSPC observation as RXJ0943.0+4701 by~\cite{Sch96a} and had a measured flux in the ROSAT $\left(0.1-2.4\ {\rm keV}\right)$ band of $f_{\rm ROSAT,PSPC}=4\times 10^{-14}\ {\rm erg\ s}^{-1}{\rm cm}^{-2}$. During a second ROSAT observation of the cluster done by~\cite{Sch98} using the HRI, the source was hardly visible with a resulting flux in the $\left(0.1-2.4\ {\rm keV}\right)$ band of $f_{\rm ROSAT,HRI}=5\times 10^{-15}\ {\rm erg\ s}^{-1}{\rm\ cm}^{-2}$, indicating that this source has a very strong variability. The source coordinates observed by PSPC are slightly offset ($\approx 38\arcsec$ south) with respect to the coordinates observed with XMM.\\
We find the galaxy Abell 0851:[SED95] 066 as a possible optical counterpart, which is a probable member galaxy. The spectral fit yields $\Gamma=1.8\pm0.1$; $f_{\rm X}(0.3-10.0)=6.8\times 10^{-14}\ {\rm erg\ s}^{-1}{\rm cm}^{-2}$; $L_{\rm X}{\rm (bol)}=1.2\times 10^{44}\ {\rm erg\ s}^{-1}$. We compute a flux in the $\left(0.1-2.4\ {\rm keV}\right)$ band of $f_{\rm XMM}=4.7\times 10^{-14}\ {\rm erg\ s}^{-1}{\rm cm}^{-2}$, which is only slightly higher than the values computed by~\cite{Sch96a} in their first X-ray observation of the cluster. Considering the changing values of the luminosity and the results of the spectral analysis, the conclusion drawn by~\cite{Sch98}, of source $\#11$ being a strong X-ray variable AGN, is, up to now, the most reliable one.
\item[ID $\#12$] No optical counterpart was found for this source. The spectral fit yields $\Gamma=1.95\pm0.2$; $f_{\rm X}(0.3-10.0)=2.0\times 10^{-14}\ {\rm erg\ s}^{-1}{\rm cm}^{-2}$.
\item[ID $\#13$] No optical counterpart. The spectral fit yields $\Gamma=1.7^{+0.2}_{-0.3}$; $f_{\rm X}(0.3-10.0)=2.5\times 10^{-14}\ {\rm erg\ s}^{-1}{\rm cm}^{-2}$.
\item[ID $\#14$] Probable member galaxy Abell 0851:[DG92] 292. The spectral fit yields $\Gamma=1.6\pm0.1$; $f_{\rm X}(0.3-10.0)=3.3\times 10^{-14}\ {\rm erg\ s}^{-1}{\rm cm}^{-2}$; $L_{\rm X}{\rm (bol)}=5.4\times 10^{43}\ {\rm erg\ s}^{-1}$.
\item[ID $\#15$] Probable member galaxy Abell 0851:[DG92] 064. Assuming a power law spectrum with photon index $\Gamma=1.7$ we obtain an unabsorbed flux of $f_{\rm X}(0.3-10.0)=1.8\times 10^{-14}\ {\rm erg\ s}^{-1}{\rm cm}^{-2}$, and a bolometric luminosity of $L_{\rm X}{\rm (bol)}=2.9\times 10^{43}\ {\rm erg\ s}^{-1}$. 
\item[ID $\#16$] No optical counterpart was found. Assuming a power law spectrum with photon index $\Gamma=1.7$: $f_{\rm X}(0.3-10.0)=3.7\times 10^{-14}\ {\rm erg\ s}^{-1}{\rm cm}^{-2}$.
\item[ID $\#17$] This source is a QSO (known as Abell 0851:[DG92] 440) lying in the field of view of the cluster and at a redshift of $z=2.055$~\citep{Dre93}. The spectral fit yields $\Gamma=1.72\pm0.05$; $f_{\rm X}(0.3-10.0)=3.1\times 10^{-14}\ {\rm erg\ s}^{-1}{\rm cm}^{-2}$; $L_{\rm X}{\rm (bol)}=2.3\times 10^{45}\ {\rm erg\ s}^{-1}$. In the ROSAT energy band the QSO has a luminosity of $L_{\rm XMM}\left(0.1-2.4\right)=8.6\times 10^{44}\ {\rm erg\ s}^{-1}$, which is slightly lower than was measured by~\cite{Sch98} ($L_{\rm ROSAT}\left(0.1-2.4\right)=1.4\times 10^{45}\ {\rm erg\ s}^{-1}$). Due to the low number of counts~\cite{Sch98} could not fit any spectral model to the data and, to compute the luminosity, they fixed the photon index at $\Gamma=2.3$. Using the same value for the photon index the two luminosities, computed by ROSAT and XMM, are perfectly consistent.
\item[ID $\#18$] Probable member galaxy Abell 0851:[DG92] 105. Assuming a power law spectrum with photon index $\Gamma=1.7$ we obtain an unabsorbed flux of $f_{\rm X}(0.3-10.0)=1.9\times 10^{-14}\ {\rm erg\ s}^{-1}{\rm cm}^{-2}$, and a bolometric luminosity of $L_{\rm X}{\rm (bol)}=3.1\times 10^{43}\ {\rm erg\ s}^{-1}$. 
\item[ID $\#19$] J0943.2+4700; has no optical counterpart. The spectral fit yields $\Gamma=1.8\pm0.1$; $f_{\rm X}(0.3-10.0)=7.8\times 10^{-14}\ {\rm erg\ s}^{-1}{\rm cm}^{-2}$.
\end{description}

\subsection{Comparison with the optical waveband}
\begin{figure*}
\centering
\includegraphics[width=17cm]{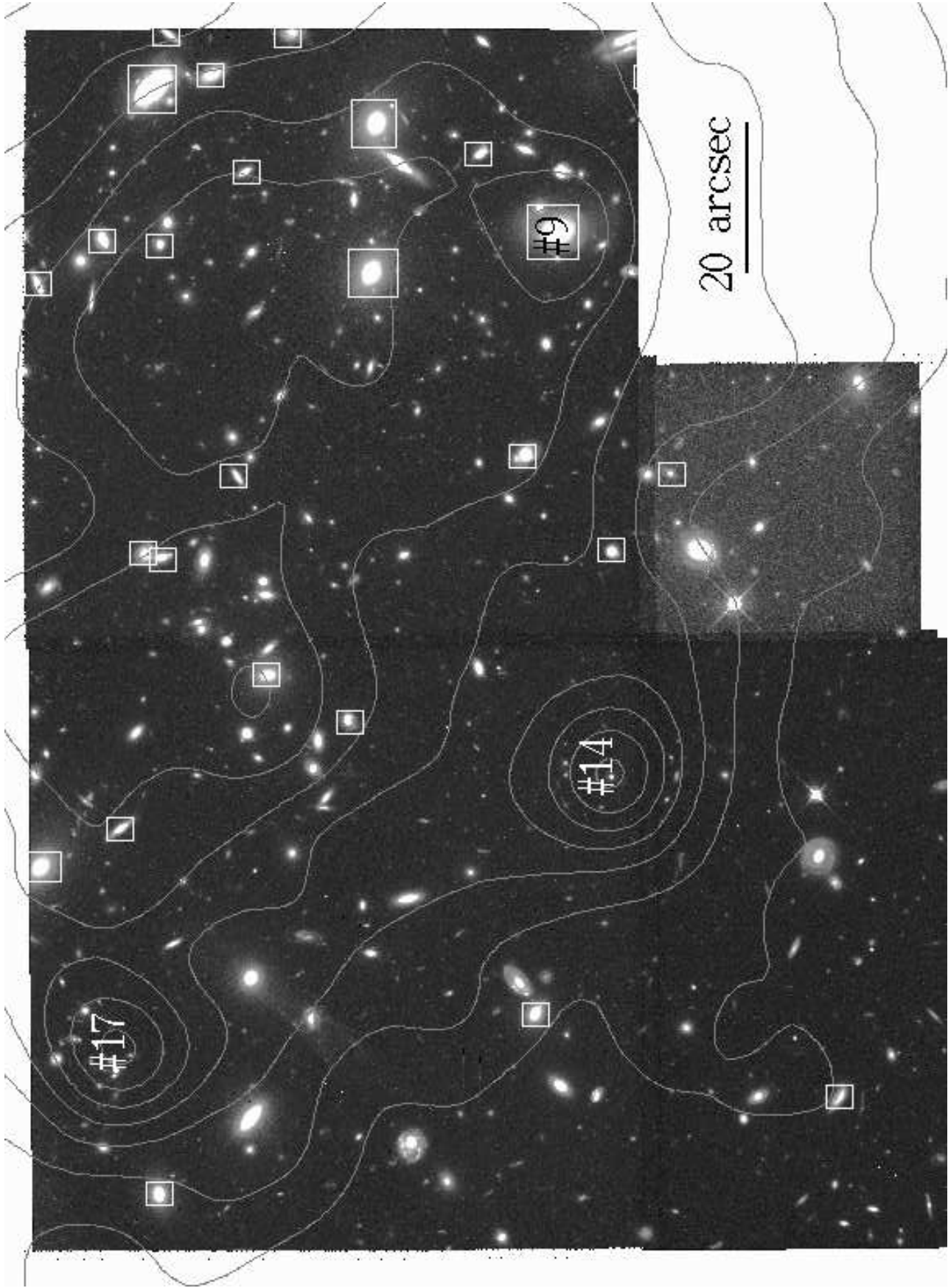}
\caption{HST image of the central region of CL~0939+4713~\citep{Bus00}. Superposed are the $\left( 0.3-10.0\ {\rm keV}\right)$ XMM X-ray contours linearly spaced from $0.24$ to $0.98$ of the maximum. White rectangles surround galaxies known to be cluster members.}
\label{fig:HST}
\end{figure*}
In Fig.~\ref{fig:HST} an HST image of the central area of the cluster is shown. On it, superposed, are the X-ray contours.\\
An extensive redshift survey of the galaxies in the field of view of CL~0939+4713 was performed by~\cite{Dre99} and at least $60$ galaxies were spectroscopically proven to belong to the cluster. Among these galaxies, those falling within the field of view of HST, can be seen in Fig.~\ref{fig:HST} surrounded by rectangles.\\
Some of these galaxies (e.g. $\#9$) are very luminous also in  X-rays but, nevertheless, the X-ray luminosity of the central region of the cluster is strongly dominated by extended hot emission from the intra-cluster gas, peaked in the two subclusters.\\
Both the galaxy number density distribution and the X-ray emission show two main peaks. The peaks of these two distributions are in similar regions. It is also interesting that the mass distribution (as computed by~\cite{Sei96} using a weak lensing analysis) has a similar bimodal structure.\\
The subcluster M1 coincides with the strongest maxima in both the galaxy number density and the mass distribution. The peaks of the galaxy density and of the mass distributions are $\approx 26\ {\rm arcsec}$ West and $\approx 15\ {\rm arcsec}$ South of the X-ray peak M1.\\
The other subcluster, M2, is located near the second maxima in the galaxy number density and in the mass distribution (their peaks are, respectively,  $\approx 16\ {\rm arcsec}$ west and $\approx 20\ {\rm arcsec}$ south-east of the X-ray one).\\

%__________________________________________________________________

\section{Spectral analysis}
For the whole spectral analysis, the ready-made on-axis response files (m1\_r5\_all\_15.rsp for MOS1, m2\_r5\_all\_15.rsp for MOS2~\footnote{Both MOS files are for patterns from $0$ to $12$.} and epn\_ff20\_sY9.rmf for pn~\footnote{pn response file is for single-pixel events only.}) were used.~\footnote{All response files can be downloaded via anonymous ftp at {\it xmm.vilspa.esa.es} in the directory /pub/ccf/constituents/extras/responses.}\\
Source and background spectra were accumulated separately for each detector. Spectra were re-binned in order to obtain $S/N > 5\ \sigma$ in each bin after background subtraction.\\
As the spectra are already corrected for vignetting, spectra of the same physical region, observed with the three different detectors, can simply be fitted simultaneously in order to maximize the signal to noise ratio.\\
All the spectral analysis was performed in the $0.3-10.0\ {\rm keV}$ band. Data  below $0.3\ {\rm keV}$ have been excluded to avoid residual calibration problems in the MOS \& pn response matrices below these energies.\\

\subsection{Results}
\begin{figure}
\centering
\includegraphics[angle=-90,width=8cm]{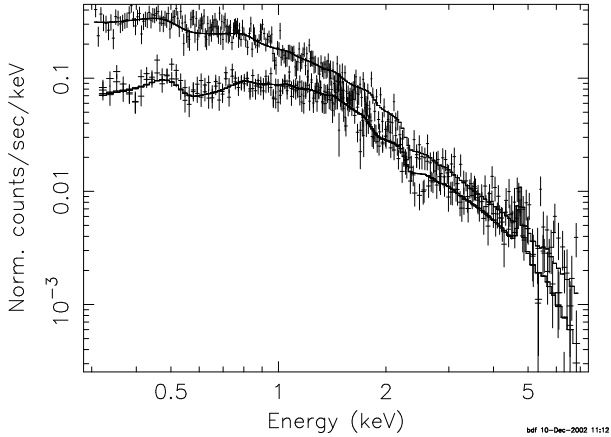}
\caption{MOS1, MOS2 \& pn spectra with MeKaL model superimposed (plotted in different shades of gray).}
\label{fig:spec_cluster}
\end{figure}
The spectra for the cluster analysis were extracted from a circular region centered at ${\rm RA}=09^{\rm h}\ 43^{\rm m}\ 01.0^{\rm s}$, ${\rm DEC}=+46\degr\ 59\arcmin\ 37.5\arcsec$ (J2000) and of radius r=$3.0\arcmin$. The emission coming from all the point sources inside the region was excluded.\\
X-ray temperatures, metal abundances, fluxes and luminosities, were computed with XSPEC fitting the data to an absorbed isothermal plasma emission code of~\cite{Kaa93} including the FeL calculations of~\cite{Lie95}~\footnote{MeKaL code from now on.}, folded through the appropriate response matrices. In Table~\ref{tab:spec_cluster} is a detailed description of the results obtained fitting the theoretical model to data from the three detectors at the same time, and for each detector independently.\\
Temperatures obtained using MOS and pn detectors independently are largely consistent. The fit results for the redshift are consistent with its spectroscopic value computed by~\cite{Dre99} ($z=0.406$); good agreement is also observed between the fit value of the hydrogen column density and its Galactic value of $n_{\rm H}=1.24\times10^{20}\ {\rm cm}^{-2}$~\citep{Dic90}. When fixing these two parameters the best fit temperature and metal abundance of the cluster based on a simultaneous fit of all (MOS+pn) the spectra (using independent normalizations for the MOS and the pn cameras) are $kT=4.5\pm 0.2\ {\rm keV}$ and $Z=0.20\pm 0.06\ Z_{\odot}$, respectively (the corresponding MOS1, MOS2 \& pn spectra with the fitted MeKaL model superimposed are plotted in Fig.~\ref{fig:spec_cluster}).\\
Using the best fit model parameters we derived, out to a radius of $210 \arcsec$, an unabsorbed flux of $f_{\rm X}(0.3-10.0)=1.3\pm 0.1\times 10^{-12}\ {\rm erg\ s}^{-1}{\rm cm}^{-2}$, corresponding to a luminosity, in the cluster rest frame, of  $L_{\rm X}(0.3-10.0)=1.1\pm 0.1\times 10^{45}\ {\rm erg\ s}^{-1}$ and a bolometric luminosity of $L_{\rm X}{\rm (bol)}=1.3\pm 0.1\times 10^{45}\ {\rm erg\ s}^{-1}$.\\

\begin{table*}[t]
\caption[]{Results of the spectral fits for the whole cluster. Column 1: line number. Column 2: detectors in use. In Cols. 3-6 the fit parameters (temperature in keV, metallicity in solar units, redshift and HI column density in $10^{20}\ {\rm cm}^{-2}$) are listed. In Col. 7 is the fit reduced $\chi^2$. }
\begin{center}
\begin{tabular}{ l l l l l l l l l }
\hline \hline
        &       Detector& Temperature&  Metallicity&    Redshift & HI column density & $\chi^2/{\rm d.o.f.}$ \cr
        &               & kT in keV&    in solar units&   & $\left [10^{20}\ {\rm cm}^{-2}\right]$&     \cr
\hline
1 &     MOS1,MOS2 \& pn &   $4.5\pm 0.2$ & $0.20\pm 0.06$&  $0.406$ (fixed) & $1.24$ (fixed) &       $1.09$\cr
2 &     MOS1,MOS2 \& pn &   $4.6\pm 0.2$ & $0.22\pm 0.06$&  $0.418^{+0.006}_{-0.010}$ & $1.24$ (fixed) &        $1.10$\cr
3 &     MOS1,MOS2 \& pn &   $4.6^{+0.4}_{-0.2}$ & $0.24^{+0.06}_{-0.07}$&  $0.41^{+0.01}_{-0.01}$ & $0.3^{+0.9}_{-0.3}$ & $1.10$\cr
4 &     MOS1,MOS2   &       $4.6\pm 0.3$ & $0.23\pm 0.08$&  $0.406$ (fixed) & $1.24$ (fixed) &       $1.11$\cr
5 &     MOS1,MOS2   &       $4.6\pm 0.3$ & $0.24\pm 0.08$&  $0.41\pm 0.01$ & $1.24$ (fixed) &        $1.11$\cr
6 &     pn      &       $4.4\pm 0.4$ & $0.16^{+0.09}_{-0.08}$&  $0.406$ (fixed) & $1.24$ (fixed) &       $1.09$\cr
7 &     pn      &       $4.5\pm 0.4$ & $0.19\pm 0.1$&  $0.43\pm 0.02$ & $1.24$ (fixed) &        $1.10$\cr
8 &     pn      &       $5.1^{+0.5}_{-0.3}$ & $0.21\pm0.1$&  $0.43\pm0.02$ & $0.2^{+1.0}_{-0.2}$ &        $1.10$\cr
\hline
\end{tabular}
\end{center}
\label{tab:spec_cluster}
\end{table*}

\subsection{Temperature map}
Since the cluster has been observed to have an irregular morphology, a deeper analysis of its internal structure and dynamical state is required in order to gain a better understanding of it.\\
The cluster was subdivided into $18$ regions; their sizes have been chosen such that each region (after background subtraction) has similar number of photon counts. We fixed the HI column density and the redshift to the average values for the whole cluster and then fitted the spectrum for each region to obtain the values of the temperature.\\
\begin{figure}
\centering
\includegraphics[width=8.8cm]{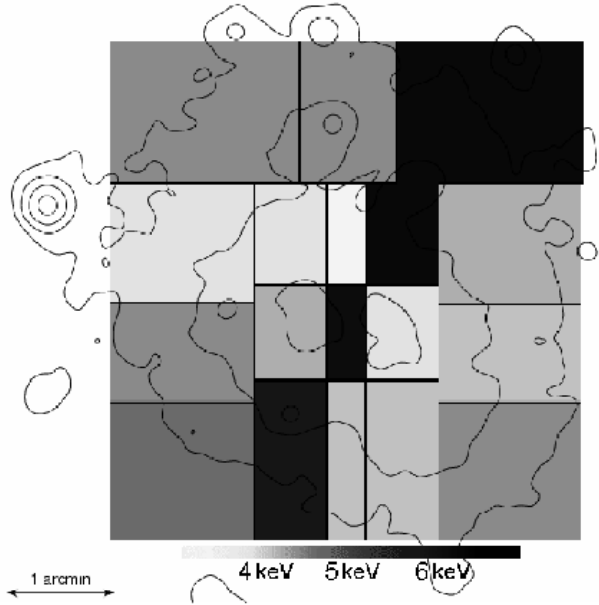}
\caption{Cluster temperature map; superposed are linearly spaced X-ray contours. Darker shades of gray indicate higher temperature. A trend of higher temperatures is observed in the central region, located between the two peaks (${\rm P}_{\rm M1}$ and ${\rm P}_{\rm M2}$) in the X-ray emission. The high temperature region also extends in the NW and SE directions.}
\label{fig:temp_map}
\end{figure}
The resulting temperature map is shown in Fig.~\ref{fig:temp_map}, where different shades of gray reflect the different temperatures of the various regions. Superposed are linearly spaced X-ray contours.\\
A clear trend of higher temperatures is observed in the region situated in the centre between the two clumps forming the core. The region of higher temperature also extends in the NW and SE directions.\\
Average temperatures were also computed from spectra extracted from the central hotter region and from the two regions surrounding it. 
All three regions have elliptical shapes and the coordinates of their centres, and the major and minor axes are: M2. RA$=9^{\rm h}\ 43^{\rm m}\ 08.11^{\rm s}$, DEC$=+47\degr\ 00\arcmin\ 03.48\arcsec$ (J2000), $a=1.5\arcmin$, $b=1.2\arcmin$, P.A.$=140\degr$; Central. RA$=9^{\rm h}\ 43^{\rm m}\ 01.58^{\rm s}$, DEC$=+46\degr\ 59\arcmin\ 18.02\arcsec$ (J2000), $a=2.1\arcmin$, $b=0.4\arcmin$, P.A.$=159\degr$; M1. RA$=9^{\rm h}\ 42^{\rm m}\ 55.50^{\rm s}$, DEC$=+46\degr\ 58\arcmin\ 35.50\arcsec$ (J2000), $a=1.4\arcmin$, $b=1.0\arcmin$, P.A.$=179\degr$. 
A MeKaL model was used; HI column density and the redshift are fixed to the average values for the whole cluster. The resulting temperatures, which clearly confirm the trend observed in the temperature map, are $kT=4.8^{+0.5}_{-0.4}\ {\rm keV}$, $kT=6.0^{+0.7}_{-0.6}\ {\rm keV}$ and $kT=4.7^{+0.4}_{-0.3}\ {\rm keV}$ (for M2, central and M1 regions, respectively). Detailed results of the fits are presented in Table~\ref{tab:spec_regions}.\\
\begin{table*}[t]
\caption[]{Results of the spectral fits for the three subregions using a MeKaL model. HI column density and redshift are fixed to the average values for the whole cluster. Column (1): region analyzed. In Cols. (2) and (3) the fit parameters (temperature in keV and metal abundance, respectively) are listed. In Col. (4) is the reduced $\chi^2$ of the fit.}
\begin{center}
\begin{tabular}{l l l l l}
\hline \hline
Region & Temperature&   Metallicity& $\chi^2/{\rm d.o.f.}$ \cr
    & kT in keV& in solar units &       \cr
\hline
M2 &    $4.8^{+0.5}_{-0.4}$ & $0.15\pm 0.11$ & $0.96$ \cr
Central &       $6.0^{+0.7}_{-0.6}$ & $0.11^{+0.18}_{-0.11}$& $1.03$ \cr
M1 &    $4.7^{+0.4}_{-0.3}$ & $0.33^{+0.14}_{-0.12}$& $1.07$ \cr
\hline
\end{tabular}
\end{center}
\label{tab:spec_regions}
\end{table*}
The temperature map in Fig.~\ref{fig:temp_map} is similar to the predictions of simulations of cluster merging, for an almost head-on encounter of two subclusters of similar mass, shortly before the merging process~\citep{Ric01,Rit02,Tak99}. Similar temperature structure has also been observed in clusters such as A399 and A401~\citep{Mar98} and A3528, A1750, and A3395~\citep{Don01}.\\
The central hotter region therefore delineates an area where the two subclusters are compressing and heating intra-cluster gas, while moving one towards the other in the beginning of a merging process. Simulations by~\cite{Sch93} show that this hotter region between the two merging structures begins to appear $\approx 0.5-1.0\ {\rm Gyr}$ before the merging. Comparing this result with simulations performed by~\cite{Tak99,Tak00} allows a determination of the dynamical state: the two subclusters in  CL~0939+4713 will collide in a few hundreds of ${\rm Myr}$.\\

\subsection{Metallicity variation}

A higher metal abundance is observed in region M1 compared to the abundance in M2 and to the abundance in the central region (see Table~\ref{tab:spec_regions}). While in M1 a metallicity of $0.33^{+0.14}_{-0.12}$ is found, the metallicity in M2 amounts only to $0.15\pm0.11$ in solar units. The errors are relatively large because only photons from small areas were used. The 90\% errors overlap slightly. But there is a clear trend that the metallicity is not constant.
The high-metallicity region M1 coincides with a strong peak in the galaxy number density distribution. It seems that the galaxies present in this region have  enriched the intra-cluster gas with heavy elements. It is interesting that the high metal content is still localised within this region despite the active dynamical state of the cluster. In order to maintain such a high metallicity during the merger process, the subcluster M1, on its way towards the subcluster M2, did obviously, not accumulate a lot of gas from the front and did not leave large quantities of its gas behind, always retaining the same gas. An additional explanation could be that the gas in this region has been metal enriched only recently, so that the motion of the gas has not had time yet to distribute the metals.\\

%------------------------------------------------------------------------------

\section{Mass analysis}

\begin{figure}
\centering
\includegraphics[width=8.8cm]{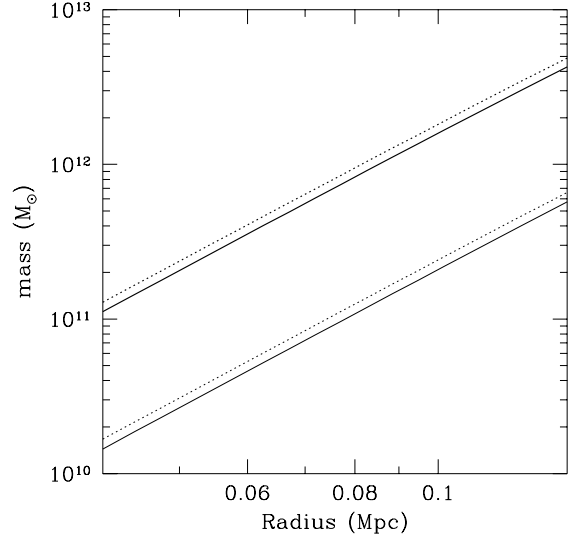}
\caption{Integrated total mass (thick lines, top) and gas mass (thin lines, bottom) profiles for the subclusters M1 (solid lines) and M2 (dotted lines).}
\label{fig:M1M2mass}
\end{figure}

The major difficulty to estimate the total mass in this cluster is its active dynamical estate. CL~0939+4713 is undergoing a major merger and the hypothesis of hydrostatic equilibrium and spherical symmetry are in principle not valid. We therefore instead assume a bimodal model where each subcluster follows an isothermal beta-model. To estimate the gas and total mass of the cluster we analyse each subcluster separately assuming for each of them hydrostatic equilibrium and spherical symmetry.\\
The integrated mass contained within a radius $r$ for each subcluster can be calculated from the equation
\begin{equation}
M(<r)=-\frac{kr}{\mu m_{\rm p}G}T_{\rm gas}\left (\frac{d\ln \rho_{\rm gas}(r)}{d\ln r}+\frac{d\ln T_{\rm gas}(r)}{d\ln r}\right)
\label{eq:mass}
\end{equation}
where $\rho_{\rm gas}$ and $T_{\rm gas}$ are the density and the temperature of the intracluster gas, and $r$,$k$,$\mu$,$m_{\rm p}$, and $G$ are the radius, the Boltzmann constant, the molecular weight, the proton mass, and the gravitational constant, respectively.\\
To estimate the gas and total mass for each subcluster, the $\beta$ parameters obtained in the fits of the profiles centred on the peaks ${\rm P}_{\rm M1}$ and ${\rm P}_{\rm M2}$ are used (see Table~\ref{tab:parameters}). We assume a constant temperature of 4.7 and 4.8 keV for the subclusters M1 and M2, respectively. 
The two subclusters are very close each other: the outer radius of $\approx22''$ (140 kpc) corresponds to half of the distance between the subclusters. Therefore we can compute the gas mass and total cluster mass out to this radius adding the masses corresponding to each subcluster. Fig.~\ref{fig:M1M2mass} shows the integrated gas and total masses out to 140 kpc for the two subclusters independently and the numerical values are given in Table~\ref{tab:M1M2mass}.\\
\begin{table}[ht]
\caption{Numerical values for the mass determination in the two subclusters M1 and M2.}
\begin{center}
\begin{tabular}{c c c}
\hline \hline
  &gas mass ($10^{12}M_{\odot}$) & total mass $M(r)$ ($10^{12}M_{\odot}$)\cr
  & within 140 kpc& within 140 kpc\cr
\hline
M1 & 0.56 &$4.2^{+0.8}_{-0.7}$\cr
M2 & 0.65 &$4.8^{+0.6}_{-0.5}$\cr
\hline
\end{tabular}
\end{center}
\label{tab:M1M2mass}
\end{table}
For an estimation of the total cluster mass out to a larger radius we use the fit centred on the subcluster M2, where the X-ray emission is higher. In Table~\ref{tab:M2mass} the gas and total masses found at two different radii are listed. At radius $r_{500}=0.85\ {\rm Mpc}$ the total cluster mass and gas mass are: $M_{\rm tot}\left(r_{500}\right) =2.3 \ \times 10^{14} \msol$ and $M_{\rm gas}\left(r_{500}\right) =0.49 \times 10^{14} \msol$. The gas mass fraction tends to increase slightly outwards ($\approx$17\% at 0.5 Mpc and $\approx$22\% at 1 Mpc).\\
The tabulated errors on the total mass come from the uncertainty in the temperature and the errors in the $\beta$ parameters. The emissivity of the gas in the XMM energy band $0.3-2.0\ {\rm keV}$ is almost constant inside the uncertainties for the estimate temperature. Therefore we can derive the gas density distribution without the uncertainty of the temperature estimate.\\
\begin{table}[ht]
\caption[]{Numerical values for the mass determination in cluster CL~0939+4713 for two specific radii; the masses are computed using the fit parameters for the profile centred on subcluster M2.}
\begin{center}
\begin{tabular}{l c c}
\hline \hline
  & mass within & mass within \cr
  & 0.5 Mpc     & 1.0 Mpc\cr
\hline
gas mass ($10^{14} M_{\odot}$) & 0.18 & 0.67\cr
total mass $M(r)$ ($10^{14} M_{\odot}$) &$1.01\pm0.3$&$2.98\pm0.7$\cr
\hline
\end{tabular}
\end{center}
\label{tab:M2mass}
\end{table}
The mass distribution of part of CL~0939+4713 has been reconstructed with a weak lensing analysis by~\cite{Sei96} and a comparison with our cluster mass estimation would be interesting. Unfortunately the restricted region within which the total mass has been calculated with the gravitational lensing analyses makes a proper comparison impossible.

%------------------------------------------------------------------------------

\section{Discussion of the merger hypothesis}
Several observed characteristics of CL~0939+4713, either in the X-rays and in the optical, point to a scenario in which the two subclusters M1 and M2, forming the cluster core, with masses of 4.2$\times10^{12} M_{\odot}$ and 4.8$\times10^{12} M_{\odot}$ respectively, are in the course of a merging process, moving one towards the other along a direction coincident with the cluster major axis. The cores of the subclusters will pass through each other in a few Myr.
\begin{description}
\item[1)] The spectral analysis of CL~0939+4713 has revealed an elongated region of heated gas located between the two X-ray peaks, and almost perpendicular to the cluster major axis. Hydrodynamic simulations~\citep{Sch93,Ric01,Rit02,Tak99} show that the gas located between two subclusters is compressed and heated during the approach of the subclusters. This hotter gas tends to expand in the direction perpendicular to the collision axis, therefore it shows an elongated shape perpendicular to the collision axis. The gas temperature and the elongation of the central region increase as the two subclusters get closer, while the bulk of the X-ray emission still comes from the cores of the two subclusters until the two cores collide. 
\item[2)] CL~0939+4713 shows an elongated X-ray morphology. Also this feature is consistent with simulations of merging clusters~\citep{Sch02}, which show that before the merger the intracluster gas is elongated along the collision axis.
\item[3)] The X-ray maxima of the two subclusters are displaced with respect to the peaks in the galaxy density distribution. In M1 the X-ray peak seems to be ahead of the galaxy concentration. Obviously, the galaxies do not trace the dark matter distribution well in this region. This is not surprising as during a merger the distributions of the different components can be quite different. 
\item[4)] CL~0939+4713 has a high velocity dispersion, $\sigma\approx 1300\ {\rm km\ s}^{-1}$~\citep{Dre92,Dre99}, indicating a relative motion of two or more components along the line of sight. The velocity dispersions of the two subclusters, computed independently, are instead slightly lower:  $\sigma\approx 1000\ {\rm km\ s}^{-1}$ and  $\sigma\approx 750\ {\rm km\ s}^{-1}$ for regions M2 and M1, respectively.
\item[5)] \cite{Dre99} observed a very high fraction of post-starburst galaxies, in CL~0939+4713, compared to other rich clusters and to field galaxies. \cite{Fuj99} have investigated the effect of ram pressure in merging clusters. They find (in contrast to what was  predicted previously by~\cite{Evr91}) that the increase in the star-formation rate caused by compression is less important than the ram-pressure stripping, which dramatically increases before a merging process. The stripping of the interstellar medium causes a decrease of the star formation rate, with a subsequent increase of the post starburst galaxy fraction.
\item[6)] \cite{And97} found that early-type galaxies in CL~0939+4713 do not show a radial symmetry, but are distributed along a privileged direction. This direction roughly coincides with the X-ray major axis and therefore with the possible collision axis.
\end{description}

%__________________________________________________________________

\section{Summary and Conclusions}
We confirm earlier results that  CL~0939+4713 is a dynamically young system ~\citep{Sch96a, Sch98}. The X-ray morphology has pronounced substructure. New findings of a hot region between the two main clumps are particularly indicative of a merger process, in which the two major subclusters will collide in a few hundreds of ${\rm Myr}$. \\
Particularly interesting are the metallicity variations found in this cluster. Metallicity variations were found before as radial gradients in clusters with cD galaxies~\citep{DeG01,Mol01,Kaa01,Gas02,Lew02}. Clusters without cD galaxies were found to show no radial gradients \citep{DeG01,Arn01b}. The problem when binning the cluster into radial bins is, that one averages over metallicity values coming from very different regions of the clusters. Therefore it is not surprising that the profiles are usually flat for non-cD clusters. We took a different approach here and found a hint that the metal abundances vary over the cluster. Of course it is quite difficult to do this in a relatively distant cluster. We therefore suggest to use this method in nearby clusters, instead of determining radial metallicity profiles.

%__________________________________________________________________

\begin{acknowledgements} 
We are very grateful to Bernd Aschenbach for making the XMM data available to us. They are part of the TS/MPE guaranteed time. We thank L.M. Buson and M. Cappellari for kindly providing the HST image ({\it grazie mille!}). A. Castillo-Morales acknowledges support by the Marie Curie Training Site grant HPMT-CT-2000-00136 by the European Commission. We also thank Dr. Phil A. James for carefully reading this paper. We also thank the referee Monique Arnaud: her suggestions and comments have greatly improved the reliability of the results obtained.

\end{acknowledgements}

\bibliographystyle{aa}
\bibliography{h3920}

\end{document}